\documentclass[twocolumn,exl]{jjap2} %% two-column layout
\usepackage{bm}

\title{Tunable Exchange Interaction in Quantum Dot Devices}

\author{Hiroyuki \textsc{Tamura}$^{1,2}$\thanks{E-mail address: tamura@nttbrl.jp}, Kenji \textsc{Shiraishi}$^{3}$ and Hideaki \textsc{Takayanagi}$^{1,2}$}

\inst{$^{1}$NTT Basic Research Laboratories, NTT Corporation, Atsugi, Kanagawa 243-0198, Japan \\
$^{2}$CREST-JST, 4-1-8 Honmachi, Kawaguchi, Saitama 331-0012, Japan.\\
$^{3}$Institute of Physics, University of Tsukuba, Tennodai, Tsukuba 305-8571, Japan.}

\abst{We theoretically discuss the Ruderman-Kittel-Kasuya-Yosida (RKKY) interaction between semiconductor quantum dots (QDs). When each QD having a local spin is coupled to the conduction electrons in semiconductors, an indirect exchange interaction, i.e., the RKKY interaction, is induced between two local spins. The RKKY interaction between QDs, which is mediated by the Fermi sea in semiconductors, is modulated by changing the Fermi energy, and the magnitude or even the sign of the exchange interaction can be tuned, which leads to a tunable magnetic transition in QD devices. We estimate the magnitude of the RKKY interaction in QDs as a function of the electron density and the inter-dot distance.}

\kword{quantum dot, RKKY interaction, ferromagnetism, spintronics, magnetic device}

\begin{document}
\maketitle

%\section{Introduction} %% No sections necessary for EXL, L, SN
Conventional magnetic devices are produced from magnetic materials. Since these magnetic materials usually include magnetic metals such as iron and manganese, the devices have to be formed separately from electric components, such as transistors, which are formed on the semiconductor LSI chip. With further integration or downsizing of the semiconductor chip, it will become desirable to make magnetic devices only from non-magnetic semiconducting materials and to integrate them on the same LSI chip. One possible way to achieve this goal is to form a lattice of QDs on periodic patterns. The kagome dot-lattice is one such artificial crystal which has a flat subband in the energy diagram and exhibits ferromagnetism\cite{Shiraishi01APL,Tamura02PRB}. Recently, a two-dimensional kagome wire network has actually been fabricated using the selective area growth technique of GaAs or InAs\cite{Mohan03APL}.

When magnetic impurities are diluted in metal, the conduction electrons are scattered by local spins, leading to the Kondo effect at low temperatures\cite{Kondo64ProgTheorPhys}. 
When a QD confines electrons which have nonzero total spin, it behaves like a local magnetic impurity which scatters the conduction electrons around it. The Kondo effect in QDs\cite{Kondo-Theory} has been widely observed in QD devices\cite{Kondo-Experiment}.
When two neighboring QDs having a local spin are coupled to the conduction electrons in semiconductors, the Ruderman-Kittel-Kasuya-Yosida (RKKY) interaction is expected to take effect between the two local spins. 
The RKKY interaction is known as an indirect exchange interaction between local magnetic impurities mediated by the Fermi sea of metals. 
The RKKY interaction in semiconductor QDs has several interesting features\cite{Tamura02Patent}: The Fermi wavelength $\lambda_F$ is typically several ten nanometers, which is very long, and it is possible to make neighboring dots that are within $\lambda_F$, where the RKKY interaction is ferromagnetic and its magnitude is quite large. Moreover, the RKKY interaction between QDs can be controlled by changing the electron density with gate electrodes, which leads to a tunable magnetic transition in QD devices.

In this Letter, we theoretically propose an exchange interaction mechanism based on the Ruderman-Kittel-Kasuya-Yosida (RKKY) interaction between semiconductor QDs. We demonstrate that one can tune the magnitude or even the sign of the exchange interaction. We estimate the magnitude of the RKKY interaction in QDs and discuss its observability in real QDs.

%%%%%%%%%%%%%%%%%%%%%%%%%%%%%%%
% \section{Theoretical Model}
%%%%%%%%%%%%%%%%%%%%%%%%%%%%%%%
We assume QDs having a local spin $\bm{S}_n\ (S=1/2,\ n=1,2,\cdots)$ are coupled to the Fermi sea, where the interaction can be described by the Kondo Hamiltonian with $s$-$d$ exchange interaction $H_{\textrm{ex}}$ given by
\begin{eqnarray}
&H_{\rm{ex}}&=\sum_{n kk'\sigma\sigma'}{J_{n}(\bm{k},\bm{k}')c_{k'\sigma '}^{\dagger}{\bm{\sigma}}_{\sigma '\sigma}c_{k\sigma}}\cdot \bm{S}_{n},
\label{Kondo Term}
\end{eqnarray}
where $J_{n}(\bm{k},\bm{k}')=-V_{n}^*(\bm{k}')V_{n}(\bm{k})U_{n}/\{(U_{n}+\epsilon_{n})\epsilon_{n}\}$, $V_n(\bm{k})$ is the coupling strength between conduction electrons and a quantum dot $n$ having a single-level energy $\epsilon_n$ and a charging energy $U_n$. 
Here, we assume that the coupling strength is given by $V_{n}(\bm{k})=V_ne^{-i\bm{k}\cdot\bm{r}_n}$. This means that the coupling is short-ranged and represented by a $\delta$-function $V_n(\bm{r})=V_n\delta(\bm{r}-\bm{r}_n)$. Later, we will modify this assumption and show the effect of finite-range coupling is small.
The indirect RKKY exchange interaction between two local spins is obtained from the second order perturbation for the ground state of conduction electrons at zero temperature as
\begin{eqnarray}
&&H_{\rm{RKKY}}=\sum_{\alpha}{\frac{
\left| \langle 0|H_{\rm{ex}} |\alpha \rangle \right|^2}
{E_0-E_{\alpha}}}=-J_{\rm{RKKY}}(\bm{S}_1\cdot \bm{S}_2),\hspace{15pt}
\label{RKKYHamiltonian}\\
&&J_{\rm{RKKY}}=\frac{8m J_{1}J_{2}}{\hbar^2 \Omega^2} \sum_{{|\bm{k}_1|<k_F}\atop {|\bm{k}_2|>k_F}}\frac{\cos\{(\bm{k}_1-\bm{k}_2)\cdot \bm{R}\}}{k_2^2-k_1^2},
\label{RKKYInteraction}
\end{eqnarray}
where $|\alpha\rangle=c_{\bm{k}_2\sigma}^{\dagger}c_{\bm{k}_1\sigma}|0\rangle$ and 
$J_{n}=-|V_{n}|^2 U_{n}/\{(U_{n}+\epsilon_{n})\epsilon_{n}\}$, $m$ is the effective mass of an electron in the Fermi sea, $\bm{R}=\bm{r}_2-\bm{r}_1$, and $\Omega$ is the volume of the conduction electron system.
Then, the RKKY interaction is obtained for three ($d=3$)\cite{Kittel68SSP}, two ($d=2$)\cite{Fischer75PRB,Beal-Monod87PRB}, and one ($d=1$)\cite{Yafet87PRB,Litvinov98PRB} dimensions as
\begin{eqnarray}
J_{\rm{RKKY}}(R)=16\pi E_F \tilde J_1 \tilde J_2 F_{d}(2k_F R),\label{RKKY}
\end{eqnarray}
where $E_F$ is the Fermi energy of the conduction electrons. The dimensionless Kondo parameter $\tilde J_n=-\Gamma_n U_{n}/[4\pi(U_{n}+\epsilon_{n})\epsilon_{n}]$ ($\Gamma_n=\pi\rho |V_n|^2$ is the elastic broadening of the energy level in  dot $n$ due to tunneling and $\rho$ is the density of states at $E_F$), and the range functions are given by $F_{3}(x)=(-x\cos x+\sin x)/4x^4$, $F_{2}(x)=-[J_0(x/2)N_0(x/2)+J_1(x/2)N_1(x/2)]$ ($J_n,N_n$: the $n$-th Bessel function of the first and second kind), $F_{1}(x)=-\textrm{si}(x)$ ($\textrm{si}(x)=-\int^{\infty}_{x}dt(\sin t/t)$: the sine-integral function).

For simplicity, we assume that all dots have the same Kondo parameter $\tilde J_n=\tilde J$. The elastic broadening $\Gamma$ sensitively depends on the details of tunneling. Here, we take the Kondo parameter $\tilde J=\Gamma /\pi U$ at $\epsilon=-U/2$, and choose a value $\tilde J=\Gamma /\pi U=0.15$, estimated from a typical Kondo device\cite{Wiel00Science}. Then, the expression of the RKKY interaction given in Eq.~(\ref{RKKY}) only depends on the Fermi energy $E_F$ and the distance $R$ between two dots. We estimate the magnitude of the RKKY interaction in Eq.~(\ref{RKKY}) assuming GaAs for the conduction region with the effective mass $m=0.067m_e$. For other materials, such as Si, we just replace the corresponding effective mass $m$ in the Fermi energy $E_F=\hbar^2k_F^2/2m$.

\begin{figure}[tb]
\begin{center}
\includegraphics[scale=0.85]{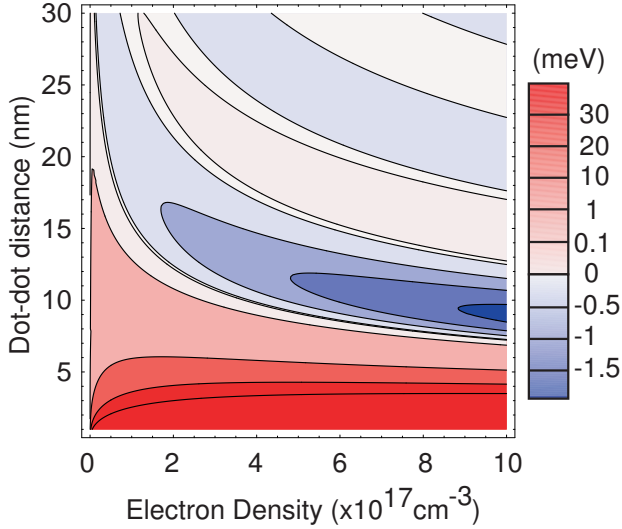}
\end{center}
\caption{The magnitude of the RKKY interaction between spins in QDs coupled to three dimensional GaAs as a function of the conduction-electron density and the inter-dot distance.}
\label{RKKY vs. Electron density in 3D}
\end{figure}

\begin{figure}[tb]
\begin{center}
\includegraphics[scale=0.85]{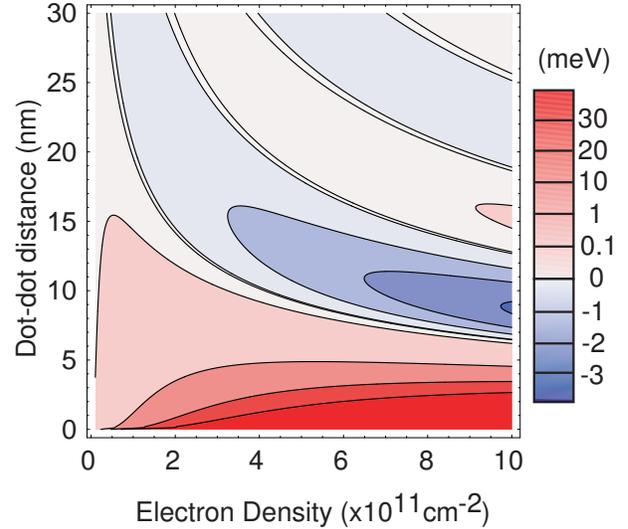}
\end{center}
\caption{The magnitude of the RKKY interaction between spins in QDs coupled to two dimensional GaAs as a function of the conduction-electron density and the inter-dot distance.}
\label{RKKY vs. Electron density in 2D}
\end{figure}

Figures~\ref{RKKY vs. Electron density in 3D} and \ref{RKKY vs. Electron density in 2D} show the electron-density dependence of the magnitude of the RKKY interaction in three and two dimensions. 
For the typical electron densities in Figs.~\ref{RKKY vs. Electron density in 3D} and \ref{RKKY vs. Electron density in 2D}, two regions, $R>1/k_F\sim 5$ nm and $R<5$ nm, have different electron-density dependence. This is due to the different asymptotic behaviors of the range function in the RKKY interaction; $F_{3}(x)\sim 1/12x\ (x\ll 1)$ and $\sim -\cos x/4x^3\ (x\gg 1)$, $F_{2}(x)\sim \ln x\ (x\ll 1)$ and $\sin x/\pi x^2\ (x\gg 1)$, and $F_{1}(x)\sim \pi/2-x\ (x\ll 1)$ and $\sim -\cos x/x\ (x\gg 1)$. In three-dimensional conduction electrons with a density region $10^{17}$~cm$^{-3}<n<10^{18}$~cm$^{-3}$, $J_{\textrm{RKKY}}^{3D}$ decays as $\cos(2k_FR)/k_F R^3$ for $R\gg 5$ nm and increases as $k_F$ for $R\ll 5$ nm with the electron density $n$, where $k_F=(3\pi^2 n)^{1/3}$. In two-dimensional conduction electrons with a density region $10^{11}$~cm$^{-2}<n<10^{12}$~cm$^{-2}$, $J_{\textrm{RKKY}}^{2D}$ oscillates as $\sin(2k_FR)/R^2$ for $R\gg 5$ nm and increases as $k_F^2\log(k_FR)$ for $R\ll 5$ nm with the electron density $n$, where $k_F=(2\pi n)^{1/2}$. It is interesting to note that, in the case of $R\gg 5$ nm in two dimensions, the oscillation amplitude of $J_{\textrm{RKKY}}^{2D}\propto \sin(2k_FR)$ does not decay with electron density $n$. 

\begin{figure}[tb]
\begin{center}
\includegraphics[scale=0.55]{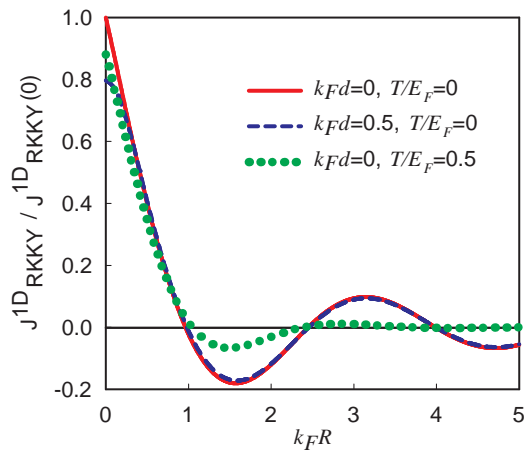}
\end{center}
\caption{Effect of a finite extent $d$ of dot and temperature $T$ on the RKKY interaction in one dimension. The RKKY interaction is normalized by the value for $R=d=T=0$.}
\label{RKKY vs. d and T in 1D}
\end{figure}

A real quantum dot is defined by a confinement potential and usually has a finite extent. To examine the effect of finite range coupling, we consider a model of potential in one dimension represented by $V_n(x)=v$ for $|x-x_n|<d/2$ and 0 for $|x-x_n|>d/2$. This potential approaches $V\delta(x-x_n)$ for $d\rightarrow 0$ by keeping $vd=V$ constant. The Fourier component of this coupling potential  is simply given by $V_n(k)=V e^{-ikx_n} \sin(kd/2)/(kd/2)$. Figure~\ref{RKKY vs. d and T in 1D} shows the calculated result for the RKKY interaction in one dimension for $k_Fd=0.5$. Although the range of each dot $d=(2k_F)^{-1}=\lambda_F/4\pi$ is considerably large, the magnitude of the RKKY interaction is only slightly reduced at $k_FR\simeq 0$. In the oscillating region $k_FR>1$, the difference is quite small.

The effect of finite temperature $T$ can be taken into account by replacing the summation to $\sum_{\mathbf{k}_1,\mathbf{k}_2}f(\epsilon_{k_1})[1-f(\epsilon_{k_2})]$ in Eq.~(\ref{RKKYInteraction}), where $f(\epsilon)=1/[1+e^{(\epsilon-E_F)/T}]$ is the Fermi-Dirac distribution function. The calculated RKKY interaction for $T=0.5E_F$ is shown in Figure~\ref{RKKY vs. d and T in 1D}. In the small-distance region $k_FR<1$, the difference is quite small, whereas, in the oscillating region $k_FR>1$, the RKKY interaction is rapidly suppressed. This behavior can be understood from Eq.~(\ref{RKKYInteraction}). At finite temperature, electron-hole pairs with different wavenumbers $\bm{k}_1$ and $\bm{k}_2$ away from $k_F$ contribute to the interaction. When the distance $\bm{R}=\bm{r}_1-\bm{r}_2$ is small, the oscillatory term $\cos[(\bm{k}_1-\bm{k}_2)\cdot \bm{R}]$ varies slowly, whereas, for large $R$, the rapidly oscillating term cancels the large contribution from excitations near $k_F$ and the RKKY interaction is significantly suppressed\cite{LeeJPC94}. 

Here, we should remark that the temperature should not exceed the charging energy $U$ in order to keep local spins well-defined in each dot. This requirement is implicitly assumed in Eq.~(\ref{Kondo Term}). However, the charging energy $U$ is typically very large in small QDs. We can estimate $U$ for a spherical QD with a diameter $L$ as $U=e^2/2\pi\epsilon L$, where $\epsilon$ is the dielectric constant. For GaAs or Si materials ($\epsilon\approx 13$), $U\approx 220$(meV)$/L$(nm). 

Quantum dots of GaAs\cite{Ueno01JJAP}, Si\cite{Oda01JPhys,ShklyaevPRB01} and Ge\cite{ShklyaevAPL02} with a diameter of 3-10 nm have already been fabricated. Dot density ranges from 10$^{12}$ to 10$^{13}$ cm$^{-2}$ and can be controlled by changing growth conditions\cite{ShklyaevPRB01,ShklyaevAPL02}. If quantum dot arrays of very small size and high density can be grown on a quantum well separated by a thin tunnel barrier, we can expect that the RKKY interaction in the arrays could be detected at very high temperatures. For example, dot arrays with a dot density of $10^{13}$ cm$^{-2}$ (or inter-dot distance of 3 nm), grown on a thin barrier layer on top of a three-dimensional quantum well of Si or GaAs, have $J_{\mathrm{RKKY}}\sim 470$ K for GaAs and 157 K for Si quantum well. The charging energy is $U\sim 850$ K for $L=3$ nm and $E_F$=825 K (GaAs) or 272 K (Si) for the conduction-electron density of $n_{\textrm{3D}}=$10$^{18}$ cm$^{-3}$. The mean level spacing $\Delta\sim E_F/N$, where $N\sim 1$ is the number of electrons in the dot, is also large and elastic cotunneling dominates inelastic processes at $T<\sqrt{U\Delta}$\cite{AverinPRL90}. Therefore, using present technology, it would be already possible to measure the effect of the RKKY interaction at room temperature. In an array of QDs with different diameters, a certain portion of dots have zero spin. These non-magnetic dots just act as normal impurities and can be ignored, if the dot density is enough high. 

The most straightforward way to detect the effect of the RKKY interaction in dot arrays is to measure the weak-localization effect in the conductivity of conduction electrons. When the electron density is modulated, the electron relaxation rates estimated from the weak-localization effect will change, depending on the spin state of QDs\cite{Wei88PRB,Vavilov03PRB}. When the dot density is low and the inter-dot distance is much larger than the Fermi wavelength, the sign of the RKKY interaction between a pair of dots can be either positive or negative. It is known that the long-range nature of the RKKY interaction with a random sign induces a spin-glass phase. It would also be interesting to study the spin-glass transition in dot-array systems by tuning the density of conduction electrons, which can be detected by measuring the magneto-transport. Another way to detect the effect of the RKKY interaction is to observe the Kondo effect in a quantum wire coupled with double QDs\cite{Tamura03PRL}. It has been shown that, when a triplet spin state is formed by the RKKY interaction, one can also tune the behavior of the Kondo screening by modulating magnetic fields.

We are grateful to W. Izumida and L. I. Glazman for valuable discussions. This work was supported by the CREST Project of the Japan Science and Technology Corporation (JST), and the NAREGI Nanoscience Project, Ministry of Education, Culture, Sports, Science and Technology, Japan. 

Note added: After completion of this work, we noticed that the RKKY interaction in QDs embedded in an Aharonov-Bohm ring is discussed by Y. Utsumi \textit{et al.}\cite{Utsumi04PRB} 
Very recently, we were aware of a paper by N. J. Craig \textit{et al}.\cite{Craig04PRB} who claim the RKKY interaction has been experimentally detected in a coupled dot system

%%%%%%%%%%%%%%%%%%%%%%%%%%%%%%%
% \section{Discussion}
%%%%%%%%%%%%%%%%%%%%%%%%%%%%%%%

%%%%%%%%%%%%%%%%%%%%%%%%%%%%%%%
% \section{Conclusions}
%%%%%%%%%%%%%%%%%%%%%%%%%%%%%%%

%%%%%%%%%%%%%%%%%%%%%%%%%%%%%%%
% \section*{Acknowledgement}
%%%%%%%%%%%%%%%%%%%%%%%%%%%%%%%


\begin{thebibliography}{99}
\bibitem{Shiraishi01APL} K. Shiraishi, H. Tamura and H. Takayanagi, Appl. Phys. Lett. \textbf{78} (2001) 3702.

\bibitem{Tamura02PRB} H. Tamura, K. Shiraishi, and H. Takayanagi, Phys. Rev. B \textbf{65} (2002) 085324.

\bibitem{Mohan03APL} P. Mohan, F. Nakajima, M. Akabori, J. Motohisa, and T. Fukui, Appl. Phys. Lett. \textbf{83} (2003) 689; P. Mohan, J. Motohisa, and T. Fukui, Appl. Phys. Lett. (in press).

\bibitem{Kondo64ProgTheorPhys} J. Kondo, Prog. Theor. Phys. \textbf{32} (1964) 37.

\bibitem{Kondo-Theory} L. I. Glazman and M. E. Raikh, JETP Lett. \textbf{47} (1988) 452; T. K. Ng and P. A. Lee, Phys. Rev. Lett. \textbf{61} (1988) 1768.

\bibitem{Kondo-Experiment} G. Goldhaber-Gordon \textit{et. al.}, Nature (London) \textbf{391} (1998) 156; S. M. Cronenwett, T. H. Oosterkamp, and L. P. Kouwenhoven, Science \textbf{281} (1998) 540; J. Schmit \textit{et al.}, Physica B \textbf{256-258} (1998) 182.

\bibitem{Tamura02Patent} H. Tamura and H. Takayanagi, Japan Patent No.2002-256036 (filed August 2002); No.2003-295541 (filed August 2003); U.S. Patent No.10/654306 (filed September 2003); E.U. and China Patents (filed September 2003).

\bibitem{Kittel68SSP} C. Kittel, in \textit{Solid State Physics}, edited by F. Seitz, D. Turnbull, and H. Ehreinreich (Academic, New York, 1968), Vol.~22, p.~1.

\bibitem{Fischer75PRB} B. Fischer and M. W. Klein, Phys. Rev. B \textbf{11} (1975) 2025.

\bibitem{Beal-Monod87PRB} M. T. Beal-Monod, Phys. Rev. B \textbf{36} (1987) 8835.

\bibitem{Yafet87PRB} Y. Yafet, Phys. Rev. B \textbf{36} (1987) 3948.

\bibitem{Litvinov98PRB} V. I. Litvinov, Phys. Rev. B \textbf{58} (1998) 3584.

\bibitem{Wiel00Science} W. G. van der Wiel, S. De Franceschi, T. Fujisawa, J. M. Elzerman, S. Tarucha, and L. P. Kouwenhoven, Science \textbf{289} (2000) 2105.

\bibitem{LeeJPC94} The suppression of the RKKY interaction in the long-distance region $k_FR\gg 1$ and at low temperature below $T/E_F< 0.05$ was also discussed by E. K. Lee, E. K. Lee, and S. Lee, J. Phys. Condens. Matter \textbf{6} (1994) 1037.

\bibitem{Ueno01JJAP} K. Ueno, K. Saiki, and A. Koma, Jpn. J. Appl. Phys. \textbf{40} (2001) 1888.

\bibitem{Oda01JPhys} S. Oda and K. Nishiguchi, J. Phys. IV (France) \textbf{11} (2001) 1065.

\bibitem{ShklyaevPRB01} A. A. Shklyaev and M. Ichikawa, Phys. Rev. B \textbf{65} (2001) 045307.

\bibitem{ShklyaevAPL02} A. A. Shklyaev and M. Ichikawa, Appl. Phys. Lett. \textbf{80} (2002) 1432.

\bibitem{AverinPRL90} D. V. Averin and Y. V. Nazarov, Phys. Rev. Lett. \textbf{65} (1990) 2446.

\bibitem{Wei88PRB} W. Wei, G. Bergmann, and R. -P. Peeters, Phys. Rev. B \textbf{38} (1988) 11751.

\bibitem{Vavilov03PRB} M. G. Vavilov, L. I. Glazman, and A. I. Larkin, Phys. Rev. B \textbf{68} (2003) 075119.

\bibitem{Tamura03PRL} H. Tamura and L. I. Glazman (to be published).

\bibitem{Utsumi04PRB} Y. Utsumi, J. Martinek, P. Bruno, and H. Imamura, condmat/0310168 (to be published in Phys. Rev. B).

\bibitem{Craig04PRB} N. J. Craig, J. M. Taylor, E. A. Lester, C. M. Marcus, M. P. Hanson, A.
C. Gossard, to be published in Science.


\end{thebibliography}
\end{document}